# The complex theory of gravitational lensing:

## Beltrami equation and cluster lensing

T. Schramm and R. Kayser

Hamburger Sternwarte, Gojenbergsweg 112, D-21029 Hamburg-Bergedorf, Germany
e-mail: tschramm@hs.uni-hamburg.de

**Abstract.** We introduce a new complex formalism to describe arclet fields in clusters of galaxies and derive the appropriate inversion techniques to find the related mass distribution of the lensing cluster. Applying the complex formalism to the statistics of lensed background sources shows that the mean of the axial ratios of locally lensed isotropically distributed elliptical sources is *not* influenced by the intrinsic ellipticity distribution, if the cluster is subcritical in that region.

## 1. Introduction

A background population of galaxies seen through a cluster shows a deformation of the morphology of each individual galaxy as imposed by gravitational lensing. If these galaxies are located near the caustics of the mapping, describing the lensing properties of the cluster, we see very elongated *giant arcs*. However all images of all galaxies, even far from the caustics, are at least slightly distorted, which are then called *arclets*. This distortion could be used to trace the potential of the cluster. If we assume the unlensed galaxies to have small, circular images, gravitational lensing stretches them to small ellipses. The axial ratio of these ellipses and their orientation could be used to find the potential, and related through Poisson's equation, the projected density distribution. In the case of infinitely small images the axial ratio is given by the ratio of the eigenvalues of the Jacobian of the lens mapping and the orientation is found by rotating the Jacobian to the diagonal form. The task is to reconstruct the potential from a measured field of axial ratios and orientations. Some authors try to fit a mass-model to the cluster to reconstruct

*Send offprint requests to*: T. Schramm

the observed arclet field, others derive a differential equation for the surface density distribution, which describes the effect approximately.

The assumption of circular sources is by no means true. The intrinsic distribution of ellipticities and orientations are not known and could only by compared statistically to fields of background galaxies which are hopefully not lensed. It must therefore be clarified in which sense a *mean* source could be accepted as circular in order to apply the techniques mentioned above.

Technically, the authors use cartesian coordinates or a vector formalism to solve the problem. We show that using a complex approach leads at least to a very elegant and compact, exact representation of the problem.

For an introduction to the theory and application of arclets see Miralda-Escudé (1991), Kaiser & Squires (1993), Breimer (1994), Broadhurst, Taylor & Peacock (1994), Carlberg, Yee & Ellington (1994), Fahlman et al. (1994), Kneib et al. (1994), Miralda-Escudé & Babul (1994), Schramm & Kayser (1994), Schneider & Seitz (1994), Smail, Ellis & Fitchett (1994a,1994b) and recently, containing most references and the basic concepts, the review of Fort & Mellier (1994).

## 2. Mathematical preliminaries

*2.1. Gravitational lensing*

The complex gravitational lens mapping is given by

$$w(x,y) = (x + \mathrm{i}y) + \left(\frac{\partial}{\partial x} + \mathrm{i}\frac{\partial}{\partial y}\right)\Phi(x,y) \qquad (1)$$

and, relating the two dimensional gravitational potential to the surface density $\sigma$,

$$-\Delta\Phi = \sigma \qquad (2)$$

$$\Phi = -\frac{1}{4\pi}\iint\left[\sigma(\xi,\eta)\log\left((\xi-x)^2 + (\eta-y)^2\right)\right]\mathrm{d}\xi\mathrm{d}\eta \qquad (3)$$

that we can express these equations in terms of *Wirtinger* derivatives (see also section "The radial case" below and compare Henrici, 1986) )

$$\frac{\partial w}{\partial z} = \frac{1}{2}\left(\frac{\partial w}{\partial x} - i\frac{\partial w}{\partial y}\right), \quad \frac{\partial w}{\partial \overline{z}} = \frac{1}{2}\left(\frac{\partial w}{\partial x} + i\frac{\partial w}{\partial y}\right) \quad . \quad (4)$$

The lens mapping is generally *not* analytic, which means that the Wirtinger derivative with respect to $\overline{z}$ is not zero. So, if we write an arbitrary complex valued function as $f(z)$ it should be implied that it could depend on $\overline{z}$ as well. We find

$$w(z) = z + 2\frac{\partial \Phi(z)}{\partial \overline{z}} \quad , \quad (5)$$

$$= z + \alpha(z) \quad , \quad (6)$$

$$-4\frac{\partial^2}{\partial z \partial \overline{z}}\Phi(z) = \sigma(z), \quad \Phi, \sigma \text{ real} \quad , \quad (7)$$

$$\Phi(z) = -\frac{1}{2\pi}\iint \sigma(\zeta) \log(|\zeta - z|) d\xi d\eta \quad (8)$$

where we introduced the complex deflection angle $\alpha$ for convenience. With Eq. (5) and Eq. (8) we find

$$w(z) = z + \frac{1}{2\pi}\iint \frac{\sigma(\zeta)}{(\overline{\zeta} - \overline{z})} d\xi d\eta \quad . \quad (9)$$

And for the derivative of Eq. 5 we find using Eq. 7

$$\frac{\partial w}{\partial z} = 1 - \frac{\sigma}{2} \quad (10)$$

and

$$\frac{\partial \alpha}{\partial z} = -\frac{\sigma}{2} \quad (11)$$

respectively.

### 2.2. The Beltrami equation

For any complex valued function $f(z)$ we find at any position $z$ an infinitesimal ellipse which is mapped onto a circle by $f$. The ellipse is characterized by the complex *dilatation* $\mu$, which is also called *Beltrami parameter*. The axial ratio of the ellipse is given by

$$\epsilon = \frac{1 - |\mu|}{1 + |\mu|} \quad (12)$$

and the argument of the major axis

$$2\varphi = \pi + \arg(\mu) \quad . \quad (13)$$

In Fig. 1 we show in the complex $\mu$-plane the corresponding ellipses.

The Beltrami parameter itself is determined in terms of Wirtinger derivatives of $f$ by the *Beltrami* equation (for details see Appendix A)

$$\frac{\partial f(z)}{\partial \overline{z}} = \mu(z)\frac{\partial f(z)}{\partial z} \quad . \quad (14)$$

**Fig. 1.** The ellipse field over the complex ($\mu = m + in$)-plane. The unit-circle is shown for convenience.

Clearly $\mu(z)$ is a complex, generally nonanalytic function over the complex plane. [1] Since the Jacobian of $f$ is

$$J = \left|\frac{\partial f(z)}{\partial z}\right|^2 - \left|\frac{\partial f(z)}{\partial \overline{z}}\right|^2 \quad (15)$$

$J = 0$ implies $|\mu| = 1$. For gravitational lensing it is interesting to note that in the matter free case ($\sigma = 0$) $\frac{\partial w}{\partial z} = 1$ and so $\mu$ is completely determined by $\frac{\partial w}{\partial \overline{z}}$. An integration, interpreted as *antiderivative*, lets us recover $w$ up to a function depending on $z$, which is just $z$ if compared with the structure of the lens mapping.

A mapping is said to be *k-quasiconformal* (q.c.) if $|\mu| \leq k < 1$ which means that there is a fixed bound on the stretching for the mapping in any given direction compared to any other direction. Lens mappings are therefore q.c. if there are no critical lines. It has been shown that there is always a unique solution for the Beltrami equation if the mapping is normalized at infinity and q.c. . Unfortunately, this applies only for marginal clusters and the proof of existence and uniqueness is hardly usable for the numerical treatment (see e.g. Carleson & Gamelin 1993, Ahlfors 1966, Henrici 1986, Imayoshi & Taniguchi 1992). In our section "Numerics" we show how to find solutions for arbitrary lens mappings. We leave it to the mathematicians (who are invited) to see if our numerical solution applies to a larger class of e. g. *Lagrange* mappings if expressed in terms of integral operators. Even existence and information we could get from arclet fields at all.

---

[1] When we discussed our paper with P. Schneider, he informed us that our $\mu$-function is the same as his $g$-function in Schneider & Seitz (1994, Eq.(2.14)), which they use in a somewhat different way.

### 2.3. Stokes's and Pompeiu's formulas for lens mappings

Pompeiu's formula is the generalization of Cauchy's formula for nonanalytical functions. Sometimes it is also called the Cauchy–Green formula, because it can be derived using Stoke's formula, which reads using Wirtinger derivatives

$$\int_{\partial R} f(z) \mathrm{d}z = 2\mathrm{i} \iint_R \frac{\partial f}{\partial \overline{z}}(z) \mathrm{d}\xi \mathrm{d}\eta \tag{16}$$

or

$$\int_{\partial R} f(z) \overline{\mathrm{d}z} = -2\mathrm{i} \iint_R \frac{\partial f}{\partial z}(z) \mathrm{d}\xi \mathrm{d}\eta \tag{17}$$

for an area $R$ with boundary $\partial R$. If applied to a function of the form $g(\zeta) := \frac{f(\zeta)}{\zeta - z}$ one finds Pompeiu's formula:

$$f(z) = \frac{1}{2\pi\mathrm{i}} \int_{\partial R} \frac{f(\zeta)}{\zeta - z} \mathrm{d}\zeta - \frac{1}{\pi} \iint_R \frac{\frac{\partial f}{\partial \overline{z}}(\zeta)}{\zeta - z} \mathrm{d}\xi \mathrm{d}\eta \quad . \tag{18}$$

The value of a function at a location inside a region $R$ bounded by $\partial R$ is given by a contour integral over the boundary plus a surface integral over the region. If we assume that

$$w(z)\big|_{|z|>R} = z - \frac{1}{\overline{z}} \tag{19}$$

which states that the lens acts as a point mass outside of the region $R$, Pompeiu's formula gives by residues for the lens mapping

$$w(z) = (1 - 1/R^2)z - \frac{1}{\pi} \iint_R \frac{\frac{\partial w}{\partial \overline{z}}(\zeta)}{\zeta - z} \mathrm{d}\xi \mathrm{d}\eta \tag{20}$$

where the $1/R^2$ term does not contribute, if in the contour integral $\zeta\overline{\zeta} = R^2 \to \infty$.

We add that if Stoke's theorem Eq. (17) is applied to the lens mapping in the form Eq. (6) we find using Eq. (11)

$$\int_{\partial R} \alpha(z) \overline{\mathrm{d}z} = \mathrm{i} \iint_R \sigma(z) \mathrm{d}\xi \mathrm{d}\eta = \mathrm{i} M(R) \tag{21}$$

which could be used for measuring the mass $M$ inside a region $R$ if the deflection angle is known on the boundary.

### 2.4. The radial case

We briefly recall how the definition of the Wirtinger derivatives is founded. Remember

$$z = x + \mathrm{i}y \quad , \tag{22}$$
$$\overline{z} = x - \mathrm{i}y \quad , \tag{23}$$
$$x = \frac{1}{2}(z + \overline{z}) \quad , \tag{24}$$
$$y = \frac{1}{2\mathrm{i}}(z - \overline{z}) \quad . \tag{25}$$

can be written as

$$\mathrm{d}f = \frac{\partial f}{\partial x}\mathrm{d}x + \frac{\partial f}{\partial y}\mathrm{d}y \quad . \tag{26}$$

If we insert the total differentials for $\mathrm{d}x, \mathrm{d}y$ as imposed by Eqs. (22–25), we find

$$\mathrm{d}f = \frac{\partial f}{\partial x}\left(\frac{\partial x}{\partial z}\mathrm{d}z + \frac{\partial x}{\partial \overline{z}}\mathrm{d}\overline{z}\right) + \frac{\partial f}{\partial y}\left(\frac{\partial y}{\partial z}\mathrm{d}z + \frac{\partial y}{\partial \overline{z}}\mathrm{d}\overline{z}\right) \tag{27}$$

$$\mathrm{d}f = \frac{1}{2}\frac{\partial f}{\partial x}(\mathrm{d}z + \mathrm{d}\overline{z}) + \frac{1}{2\mathrm{i}}\frac{\partial f}{\partial y}(\mathrm{d}z - \mathrm{d}\overline{z}) \tag{28}$$

which leads to the definition given in Eq. (4). We apply this to the polar description of complex numbers

$$z = r\mathrm{e}^{\mathrm{i}\varphi} \quad , \tag{29}$$
$$\overline{z} = r\mathrm{e}^{-\mathrm{i}\varphi} \quad , \tag{30}$$
$$r = \sqrt{z\overline{z}} \quad , \tag{31}$$
$$\varphi = \frac{1}{2\mathrm{i}}\log\frac{z}{\overline{z}} \quad . \tag{32}$$

The total derivative is

$$\mathrm{d}f = \frac{\partial f}{\partial r}\left(\frac{\partial r}{\partial z}\mathrm{d}z + \frac{\partial r}{\partial \overline{z}}\mathrm{d}\overline{z}\right) + \frac{\partial f}{\partial \varphi}\left(\frac{\partial \varphi}{\partial z}\mathrm{d}z + \frac{\partial \varphi}{\partial \overline{z}}\mathrm{d}\overline{z}\right) \tag{33}$$

$$= \frac{1}{2}\frac{\partial f}{\partial r}\left(\mathrm{e}^{-\mathrm{i}\varphi}\mathrm{d}z + \mathrm{e}^{\mathrm{i}\varphi}\mathrm{d}\overline{z}\right) \tag{34}$$

$$+ \frac{1}{2\mathrm{i}r}\frac{\partial f}{\partial \varphi}\left(\mathrm{e}^{-\mathrm{i}\varphi}\mathrm{d}z - \mathrm{e}^{\mathrm{i}\varphi}\mathrm{d}\overline{z}\right) \tag{35}$$

which imposes the definition

$$\frac{\partial f}{\partial z} = \frac{1}{2}\left(\frac{\partial f}{\partial r} - \frac{\mathrm{i}}{r}\frac{\partial f}{\partial \varphi}\right)\mathrm{e}^{-\mathrm{i}\varphi} \quad , \tag{36}$$

$$\frac{\partial f}{\partial \overline{z}} = \frac{1}{2}\left(\frac{\partial f}{\partial r} + \frac{\mathrm{i}}{r}\frac{\partial f}{\partial \varphi}\right)\mathrm{e}^{\mathrm{i}\varphi} \quad . \tag{37}$$

A (lens) mapping with radial symmetry would read

$$w(r,\varphi) = R(r)\mathrm{e}^{\mathrm{i}\varphi} \quad . \tag{38}$$

For this mapping the polar Wirtinger derivatives read simply

$$\frac{\partial w}{\partial z} = \frac{1}{2}\left(\frac{\partial R}{\partial r} + \frac{R}{r}\right) \quad , \tag{39}$$

$$\frac{\partial w}{\partial \overline{z}} = \frac{1}{2}\left(\frac{\partial R}{\partial r} - \frac{R}{r}\right)\mathrm{e}^{2\mathrm{i}\varphi} \quad . \tag{40}$$

It is convenient to set

$$\mu = |\mu|\mathrm{e}^{2\mathrm{i}\varphi} \quad , \tag{41}$$

so we find for the Beltrami equation

$$\frac{\partial R}{\partial r} - \frac{R}{r} = |\mu|\left(\frac{\partial R}{\partial r} + \frac{R}{r}\right) \quad . \tag{42}$$

find an ordinary differential equation

$$\frac{\partial R}{\partial r} - \frac{1}{\epsilon(r)}\frac{R}{r} = 0 \qquad (43)$$

which is easily solved

$$R = C \exp\left(\int \frac{\mathrm{d}r}{r\epsilon(r)}\right) \quad . \qquad (44)$$

Measuring the variation of the axial ratio leads to the radial lens equation. With Eq. (10) and the radial Beltrami equation (42) we find for the density

$$\sigma = 2 - \left(1 + \frac{1}{\epsilon}\right)\frac{R}{r} \quad . \qquad (45)$$

These results have been found using the normal vector formalism by Schramm & Kayser (1994) (see also Schneider & Seitz, 1994).

*2.5. Some instructive examples*

2.5.1. Point mass

Let us take a pure point mass acting as a lens. The lens mapping reads

$$w(z) = z - \frac{1}{\bar{z}} = \left(r - \frac{1}{r}\right)e^{\mathrm{i}\varphi} \quad . \qquad (46)$$

The ellipse field is given by the Beltrami parameter

$$\mu = \frac{1}{\bar{z}^2} = \frac{1}{r^2}e^{2\mathrm{i}\varphi} \qquad (47)$$

and the corresponding axial ratios by

$$\epsilon = \frac{|z|^2 - 1}{|z|^2 + 1} = \frac{r^2 - 1}{r^2 + 1} \quad . \qquad (48)$$

The ellipse field is given in Fig. 2. The corresponding variation of $|\mu|$ and $\epsilon$, respectively, is shown in Fig. 3.

2.5.2. Point mass and shear

We loose the radial symmetry if a shear is introduced

$$w(z) = z + \gamma\bar{z} - \frac{1}{\bar{z}} \quad . \qquad (49)$$

The corresponding ellipse field is given by the Beltrami parameter

$$\mu = \frac{1}{\bar{z}^2} + \gamma \qquad (50)$$

and the corresponding axial ratios

$$\epsilon = \frac{|z|^2 - |1 + \gamma\bar{z}^2|}{|z|^2 + |1 + \gamma\bar{z}^2|} \quad . \qquad (51)$$

The ellipse field is given in Fig. 4

**Fig. 2.** The ellipse field due to a point mass. Note that the large principal axes are fixed. For convenience we also show the critical curve.

2.5.3. Isothermal sphere

An isothermal sphere is given by a density distribution which varies as

$$\sigma = \sigma_0 \frac{1}{|z|} \quad . \qquad (52)$$

Our lens equation then reads

$$w(z) = z - \sigma_0 \frac{z}{|z|} \qquad (53)$$

and for the complex dilatation we find

$$\mu = \frac{\sigma_0 z}{\bar{z}(2|z| - \sigma_0)} \quad . \qquad (54)$$

which could, of course, be derived with the radial approach given above. We show the ellipse field in Fig. 5 and the related variation of $\epsilon$ and $|\mu|$ in Fig. 6.

## 3. Numerics

Our intention is to introduce the Beltrami approach to cluster lensing and not to discuss detailed applications. However, we briefly touch some possibilities how to solve the Beltrami equation for lensing problems. Applications will be published in a subsequent paper.

The Wirtinger derivatives are well defined as derivatives in the sense of distributions. However, it is hardly possible to give them a constructive numerical sense without casting them into their components.

As typical in potential theory we have *isolated* or *boundary value* problems. Even if we are able to define a homogeneous $\mu$-field over a region (ccd frame) for the latter case we have to guess the values of the mapping on the boundary to find a unique solution of the Beltrami equation. For this purpose we decompose the Beltrami equation into terms containing normal partial derivatives.

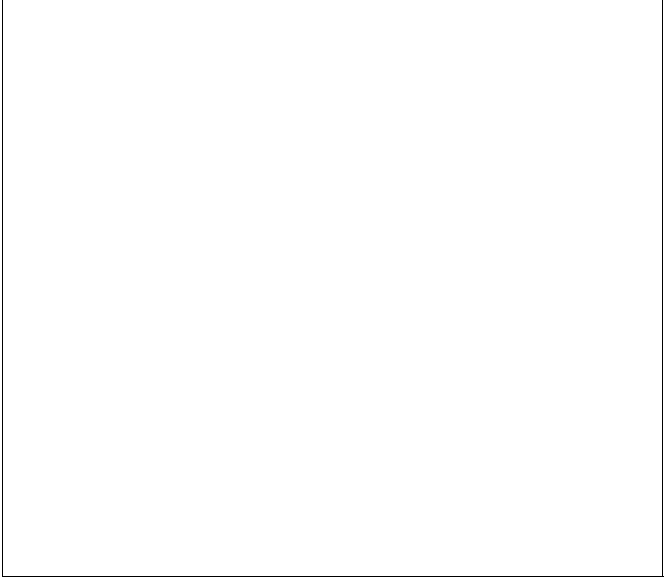

**Fig. 3.** Norm of $\mu$ and $\epsilon$ for the point mass lens. The critical line is denoted by a circle on the curve for $\mu$. The related axial ratio is zero at that point.

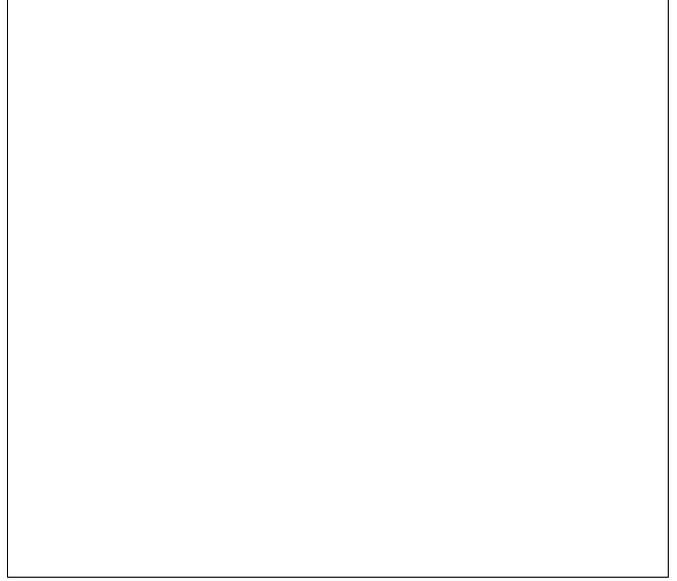

**Fig. 4.** The ellipse field due to a point mass and an additional shear. The critical curve is also shown.

$$(1-\mu)\frac{\partial}{\partial x}w(x,y) + i(1+\mu)\frac{\partial}{\partial y}w(x,y) \quad . \tag{55}$$

Introducing the complex function

$$\mathcal{E} = \frac{1-\mu}{1+\mu} \tag{56}$$

the equation reads quite simply

$$\frac{\partial w(x,y)}{\partial x} + \frac{i}{\mathcal{E}(x,y)}\frac{\partial w(x,y)}{\partial y} = 0 \quad . \tag{57}$$

This equation can be descretised and solved as a boundary value problem involving all the problems of numerical solutions of first order differential equations. We do not follow this track here but we assume that we are so lucky to have all the contributing material *in* our region with defined $\mu$-field. So we can treat our problem as *isolated*. This includes the case where the contributions of the outer parts cancel, as it is the case for radial or homoeoidal (see e.g. Schramm 1994 for details) mass distributions: If we apply the Beltrami equation to the lens mapping Eq. (9) we find:

$$\frac{1}{2\pi}\iint \frac{\sigma(\zeta)}{(\overline{\zeta}-\overline{z})^2}\mathrm{d}\xi\mathrm{d}\eta = \mu\left(1-\frac{\sigma}{2}\right) \quad . \tag{58}$$

If we now assume a grid of size $(n \times m)$ representing an area $(x_2 - x_1) \times (y_2 - y_1)$ in the lens plane with pixel size $\mathrm{d}x = (x_2 - x_1)/n$ and $\mathrm{d}y = (y_2 - y_1)/m$. We find the center coordinate $z_k = (x_i + iy_j)$ of each pixel $(i,j)$ by

$$x_i = x_1 + i\mathrm{d}x - \frac{\mathrm{d}x}{2} \quad , \tag{59}$$

$$y_j = y_1 + j\mathrm{d}y - \frac{\mathrm{d}y}{2} \quad . \tag{60}$$

We introduce the numbering scheme

$$k = (j-1)n + i \tag{61}$$

with the inversion

$$i = \mathrm{mod}((k-1)/n) + 1 \quad , \tag{62}$$

$$j = \mathrm{int}((k-1)/n) + 1 \quad . \tag{63}$$

For convenience we define a distance (matrix) function $\mathsf{D}$ on the grid

$$\mathsf{D}_{k,l} = \frac{1}{2}\begin{cases} \frac{\mathrm{d}x\mathrm{d}y}{\pi(\overline{z}_l - \overline{z}_k)^2} & \text{if } k \neq l \\ \mu_k & \text{otherwise} \end{cases} \tag{64}$$

which applies to the Beltrami equation if the integral is approximated by a sum over all pixels with constant density. We find for each pixel $k$:

$$\frac{\mathrm{d}x\mathrm{d}y}{2\pi}\sum_{l\neq k}^{nm}\frac{\sigma_l}{(\overline{z}_l - \overline{z}_k)^2} = \mu_k\left(1 - \frac{\sigma_k}{2}\right) \tag{65}$$

which, after introduction of vectors $\boldsymbol{\sigma} = (\sigma_1 \ldots \sigma_{nm})$ and $\boldsymbol{\mu} = (\mu_1 \ldots \mu_{nm})$ can be written as matrix equation:

$$\mathsf{D}\boldsymbol{\sigma} = \boldsymbol{\mu} \tag{66}$$

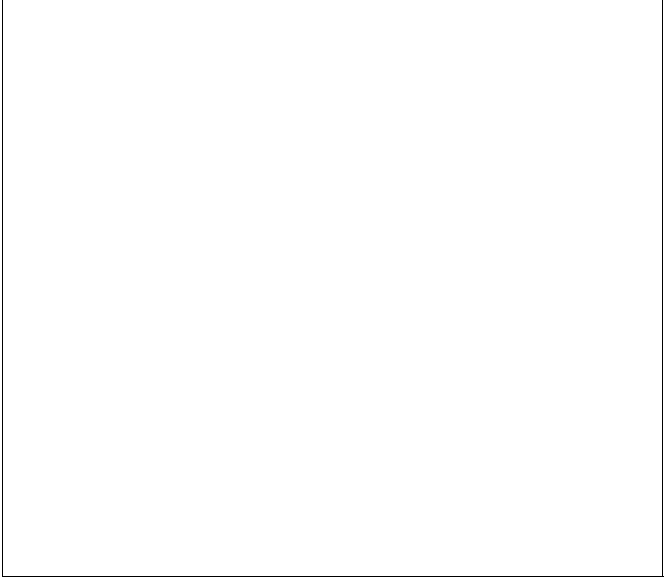

**Fig. 5.** Ellipse field of the lens as given by an isothermal sphere with $\sigma_0 = 1$. Note the tangential and radial stretching of the ellipses near the critical line and the center.

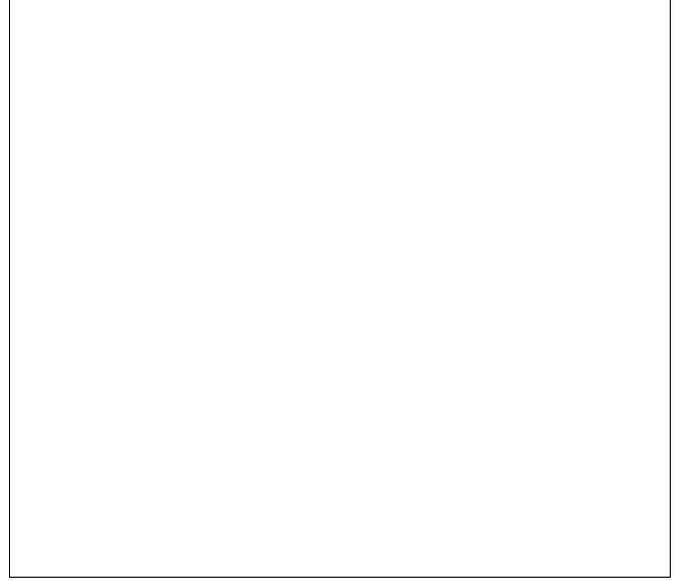

**Fig. 6.** The norm of the complex dilatation $\mu$ and the corresponding axial ratio $\varepsilon$. Note the pole of $|\mu|$ where the ellipses turn from radial to tangential stretching.

with obvious solution

$$\boldsymbol{\sigma} = \mathsf{D}^{-1}\boldsymbol{\mu} \quad . \tag{67}$$

Note that adding a sheet of constant density $\kappa$ does not alter the $\mu$-field if every mass element is scaled by a factor $1 + \kappa$ (magnification transformation). However, if we are able to observe arclets out to regions where we are sure that the local density of the cluster is zero, we can uniquely determine the total mass up to the mass contribution of the global cosmological background.

## 4. Seeing

To see the influence of the atmospheric seeing we inspect again our definition of the Beltrami parameter Eq. (12). If we assume the axis of the arclet to be $a, b$ we would find without seeing

$$|\mu| = \frac{a - b}{a + b} \quad . \tag{68}$$

With seeing $s$ the observed axis $a', b'$ are $a' = \sqrt{a^2 + s^2}$ and $b' = \sqrt{b^2 + s^2}$, respectively, if we assume a gaussian surface brightness distribution of the source. We find therefore

$$|\mu| = \frac{\sqrt{a'^2 - s^2} - \sqrt{b'^2 - s^2}}{\sqrt{a'^2 - s^2} + \sqrt{b'^2 - s^2}} \quad . \tag{69}$$

We see that we can recover the $\mu$-parameter if the seeing is known. However, we see also that we run into problems if the seeing is very bad, since the expression for $\mu$ gets indefinite.

## 5. Statistics

Observed arclets are, as mentioned above, in general not images of circular sources. Additionally, the sources are distributed over a redshift range. Although the redshift distribution can be accounted for in the Beltrami equation we here assume for simplicity that all sources are at the same redshift.

However, we can also describe the situation including intrinsic ellipticities using the Beltrami equation if a composed mapping $h$ is introduced. $h$ is a composition of two mappings w and $\omega$. The mapping w maps the observed elliptical arclets to the unknown *real* elliptical source and $\omega$ maps this source to a circle. (See Fig. 7).

The corresponding Beltrami parameters are $\mu_{\mathrm{w}}$ and $\mu_\omega$ respectively.

$$h(z) = \omega(\mathrm{w}(z)) \tag{70}$$

which all depend on complex conjugated variables as well, we find therefore for the total differential

$$\begin{aligned}
\mathrm{d}h &= \frac{\partial \omega}{\partial \mathrm{w}}\mathrm{d}\mathrm{w} + \frac{\partial \omega}{\partial \overline{\mathrm{w}}}\mathrm{d}\overline{\mathrm{w}} \quad , \tag{71}\\
&= \frac{\partial \omega}{\partial \mathrm{w}}\left(\frac{\partial \mathrm{w}}{\partial z}\mathrm{d}z + \frac{\partial \mathrm{w}}{\partial \overline{z}}\mathrm{d}\overline{z}\right) + \frac{\partial \omega}{\partial \overline{\mathrm{w}}}\left(\frac{\partial \overline{\mathrm{w}}}{\partial z}\mathrm{d}z + \frac{\partial \overline{\mathrm{w}}}{\partial \overline{z}}\mathrm{d}\overline{z}\right) \tag{72}
\end{aligned}$$

For the Wirtinger derivatives we find

$$\begin{aligned}
\frac{\partial h}{\partial z} &= \frac{\partial \omega}{\partial \mathrm{w}}\frac{\partial \mathrm{w}}{\partial z} + \frac{\partial \omega}{\partial \overline{\mathrm{w}}}\frac{\partial \overline{\mathrm{w}}}{\partial z} \quad , \tag{73}\\
\frac{\partial h}{\partial \overline{z}} &= \frac{\partial \omega}{\partial \mathrm{w}}\frac{\partial \mathrm{w}}{\partial \overline{z}} + \frac{\partial \omega}{\partial \overline{\mathrm{w}}}\frac{\partial \overline{\mathrm{w}}}{\partial \overline{z}} \tag{74}
\end{aligned}$$

Fig. 7. The composed mapping $h$ maps observed elliptical arclets to circles. It is composed from the lens mapping w that maps the observed arclet to the real elliptical source and an artificial $\omega$ mapping that maps the real elliptical source to a circle.

and so for the Beltrami parameter

$$\mu_h = \frac{\frac{\partial \omega}{\partial w}\frac{\partial w}{\partial \overline{z}} + \frac{\partial \omega}{\partial \overline{w}}\frac{\partial \overline{w}}{\partial \overline{z}}}{\frac{\partial \omega}{\partial w}\frac{\partial w}{\partial z} + \frac{\partial \omega}{\partial \overline{w}}\frac{\partial \overline{w}}{\partial z}} \quad , \tag{75}$$

$$= \frac{\frac{\partial w}{\partial \overline{z}} + \mu_\omega \frac{\partial \overline{w}}{\partial \overline{z}}}{\frac{\partial w}{\partial z} + \mu_\omega \frac{\partial \overline{w}}{\partial z}} \quad . \tag{76}$$

For simplicity we use indices in parentheses to denote derivatives. The last equation then reads

$$\mu_h = \frac{w_{(\overline{z})} + \mu_\omega \overline{w}_{(\overline{z})}}{w_{(z)} + \mu_\omega \overline{w}_{(z)}} \quad . \tag{77}$$

If we suppose now that we are in the linear regime of the mapping, we can always write

$$\omega = w + \mu_\omega \overline{w}, \quad 0 \leq |\mu_\omega| \leq 1 \quad , \tag{78}$$
$$w = (1-\sigma)z + \gamma \overline{z} \quad . \tag{79}$$

Note that we do not have to restrict the lens equation to the case of real $\gamma$. However, the only difference would be a constant phase in the appropriate Beltrami parameter, which could also be attached afterwards.

Applying Eq.(77) we find

$$\mu_w = \frac{\gamma}{1-\sigma}, \quad \text{real !} \tag{80}$$

$$\mu_h = \frac{\gamma + \mu_\omega(1-\sigma)}{1-\sigma + \gamma\mu_\omega} \quad , \tag{81}$$

$$= \frac{\mu_w + \mu_\omega}{1 + \mu_w \mu_\omega} \quad \text{and the inverse} \tag{82}$$

$$\mu_\omega = \frac{}{\mu_w \mu_h - 1} \quad . \tag{83}$$

These equations describe the transformation of the Beltrami parameter under local lensing, given by $\mu_w$. We are now interested to see how a local distribution $\Phi$ of $\mu_\omega$ transforms under local lensing. This means how the statistical properties of a field of elliptical sources alter if mapped by a lens locally described by constant density and shear:

$$\Phi(\mu_h) |\mathrm{d}\mu_h| = \Phi(\mu_\omega) |\mathrm{d}\mu_\omega| \quad , \tag{84}$$
$$\Phi(\mu_h) = J_\omega \Phi(\mu_\omega) \tag{85}$$

where we have to insert for $J_\omega$ the Jacobian of the mapping for $\mu_\omega$

$$J_\omega = \left| \frac{\partial(m_\omega, n_\omega)}{\partial(m_h, n_h)} \right|, \quad \mu = m + \mathrm{i}n \quad , \tag{86}$$

$$= \left| \frac{\partial \mu_\omega}{\partial \mu_h} \right|^2 - \left| \frac{\partial \mu_\omega}{\partial \overline{\mu}_h} \right|^2 \quad , \tag{87}$$

$$= \left| \frac{\partial \mu_\omega}{\partial \mu_h} \right|^2 \quad , \tag{88}$$

$$= \left| \frac{1 - \mu_w^2}{(1 + \mu_h \mu_w)^2} \right|^2 \quad . \tag{89}$$

We thus find for $\Phi$

$$\Phi(\mu_h) = \left| \frac{1 - \mu_w^2}{(1 + \mu_h \mu_w)^2} \right|^2 \Phi(\mu_\omega(\mu_h)) \quad . \tag{90}$$

Assuming now isotropy for the original sources, we find

$$\Phi(\mu_h) = \frac{1}{2\pi} \left| \frac{1 - \mu_w^2}{(1 + \mu_h \mu_w)^2} \right|^2 \Phi(|\mu_\omega(\mu_h)|) \quad . \tag{91}$$

The normalization for polar coordinates reads as follows:

$$1 = \int_0^\infty \int_0^{2\pi} \Phi(|\mu|, \varphi)|\mu| \, \mathrm{d}|\mu|\mathrm{d}\varphi \tag{92}$$

which is for uncorrelated $|\mu|$ and $\varphi$

$$1 = \int_0^\infty \Phi_{|\mu|}(|\mu|)|\mu|\mathrm{d}|\mu| \int_0^{2\pi} \Phi_\varphi(\varphi) \, \mathrm{d}\varphi \quad . \tag{93}$$

For an isotropical distribution

$$\Phi_\varphi = \frac{1}{2\pi} \tag{94}$$

and for the radial part

$$1 = \int_0^\infty \Phi_{|\mu|}(|\mu|)|\mu| \, \mathrm{d}|\mu| \quad . \tag{95}$$

$$\langle \mu_\omega \rangle = \int_0^{\mu_\omega^{\max}} \int_0^{2\pi} |\mu_\omega| \, e^{i\varphi_\omega} \, \Phi(\mu_\omega) \, |\mu_\omega| \, d|\mu_\omega| \, d\varphi_\omega, \quad (96)$$

$$\mu_\omega = |\mu_\omega| e^{i\varphi_\omega}$$

which is in case of isotropy just

$$\langle \mu_\omega \rangle = 0 \quad (97)$$

because of the angle integration. The image $\mu_h(\langle \mu_\omega \rangle)$ is therefore

$$\mu_h(\langle \mu_\omega \rangle) = \frac{\mu_w + \langle \mu_\omega \rangle}{1 + \mu_w \langle \mu_\omega \rangle} = \mu_w \quad (98)$$

as expected. The mean of $\mu_h$ is

$$\langle \mu_h \rangle = \iint \mu_h \left| \frac{\partial(m_\omega, n_\omega)}{\partial(m_h, n_h)} \right| \Phi(\mu_\omega(\mu_h)) dm_h dn_h \quad , \quad (99)$$

$$= \iint \mu_h \Phi(\mu_\omega) dm_\omega dn_\omega \quad . \quad (100)$$

This yields for isotropy

$$\langle \mu_h \rangle = \frac{1}{2\pi} \int_0^{\mu_\omega^{\max}} \int_0^{2\pi} \frac{\mu_w + \mu_\omega}{\mu_w \mu_\omega + 1} \Phi(|\mu_\omega|) |\mu_\omega| \, d|\mu_\omega| d\varphi_\omega \quad (101)$$

$$= \frac{1}{2\pi} \int_0^{\mu_\omega^{\max}} \left[ \int_0^{2\pi} \frac{\mu_w + \mu_\omega}{\mu_w \mu_\omega + 1} d\varphi_\omega \right] \Phi(|\mu_\omega|) |\mu_\omega| \, d|\mu_\omega|. \quad (102)$$

The integral in square brackets could be evaluated if cast into real and imaginary part. The important fact to note is that its result depends on $\text{sign}((\mu_w \mu_\omega - 1)/(1 + \mu_w \mu_\omega))$. **We** use the theory of residues and cast the integral into a contour integral around the origin:

$$d\varphi_\omega = \frac{d\mu_\omega}{i\mu_\omega} \quad , \quad (103)$$

$$\int_0^{2\pi} \frac{\mu_w + \mu_\omega}{\mu_w \mu_\omega + 1} d\varphi_\omega = \frac{1}{i} \oint \frac{\mu_w + \mu_\omega}{\mu_\omega (\mu_w \mu_\omega + 1)} d\mu_\omega \quad , \quad (104)$$

$$= 2\pi \begin{cases} \mu_w & \text{if } |\mu_\omega| < \frac{1}{\mu_w} \\ \mu_w - \frac{\mu_w^2 - 1}{\mu_w} = \frac{1}{\mu_w} & \text{if } |\mu_\omega| > \frac{1}{\mu_w} \end{cases} \quad . \quad (105)$$

We therefore find for $\langle \mu_h \rangle$ using the normalization Eq. (95)

$$\langle \mu_h \rangle = \mu_w \int_0^{\frac{1}{\mu_w}} |\mu_\omega| \Phi(|\mu_\omega|) \, d|\mu_\omega| + \quad (106)$$

$$\frac{1}{\mu_w} \int_{\frac{1}{\mu_w}}^{\infty} |\mu_\omega| \Phi(|\mu_\omega|) \, d|\mu_\omega| \quad , \quad (107)$$

$$= \left( \mu_w - \frac{1}{\mu_w} \right) \int_0^{\frac{1}{\mu_w}} |\mu_\omega| \Phi(|\mu_\omega|) \, d|\mu_\omega| + \frac{1}{\mu_w} \quad (108)$$

$$= \mu_w \quad \text{if } \Phi(|\mu_\omega|) = 0 \text{ for } |\mu_\omega| > \frac{1}{\mu_w} \quad . \quad (109)$$

Since $|\mu_\omega| \leq 1$ the last condition is fulfilled for $|\mu_w| \leq 1$, which is valid for a region where the lens mapping is orientation preserving (quasi conformal). The bad news is mation, the mean eccentricity of the observed arclets is influenced by the distribution of the intrinsic eccentricities. The good news is that we can conclude the intrinsic distribution from regions where the mapping is surely quasi conformal.

*Acknowledgements.* This work was supported by the Deutsche Forschungsgemeinschaft under Schr. 417-1. We thank U. Borgeest, A. Dent, C. Dyer, P. Helbig, B. Neindorf, W. Petersen, O. Riemenschneider, S. Refsdal, P. Schneider and I. Suisalu for interesting discussions. We also thank the companies Commodore and Scientific Computers for hard- and software support.

## A. Appendix A: Why complex?

The complex notation is typically only used for mappings which are analytic. However, even for nonanalytical functions there could be a remaining structure, which makes complex functions easier to use than the 2D-vector analogue. This has e.g. been shown for the calculation of caustics (Witt 1990) and for the computation of elliptical lenses (Kassiola & Kovner 1993, Schramm 1994). We show now why this is also the case for cluster lensing (compare Ahlfors 1966, p4).

Suppose an arbitrary local (linear) map given by an expansion of $w = w(z)$, with $w = u + iv$ and $z = x + iy$

$$u = u_{(x)} x + u_{(y)} y \quad , \quad (A1)$$
$$v = v_{(x)} x + v_{(y)} y \quad (A2)$$

where, as above, subscripts in parentheses denote differentiation. A local ellipse is then mapped onto a circle as given by the squared deviation function $SDF$ (compare Schramm & Kayser, 1987)

$$u^2 + v^2 = \left( u_{(x)} x + u_{(y)} y \right)^2 + \left( v_{(x)} x + v_{(y)} y \right)^2 = \text{const.} \quad (A3)$$

which reads in classical notation

$$u^2 + v^2 = E x^2 + 2F xy + G y^2 = \text{const.} \quad (A4)$$

with

$$E = u_{(x)}^2 + v_{(x)}^2 \quad , \quad (A5)$$
$$F = u_{(x)} u_{(y)} + v_{(x)} v_{(y)} \quad , \quad (A6)$$
$$G = u_{(y)}^2 + v_{(y)}^2 \quad . \quad (A7)$$

The axial ratio of the ellipse is then given by the ratio of the Eigenvalues $\lambda_1, \lambda_2$ determined by

$$\begin{vmatrix} E - \lambda & F \\ F & G - \lambda \end{vmatrix} = 0 \quad . \quad (A8)$$

We find

$$\lambda_1, \lambda_2 = \frac{E + G \pm \sqrt{(E - G)^2 + 4 F^2}}{2} \quad (A9)$$

$$\epsilon = \sqrt{\frac{\lambda_2}{\lambda_1}} = \frac{2\sqrt{EG - F^2}}{E + G + \sqrt{(E-G)^2 + 4F^2}} \quad . \tag{A10}$$

For the angle of *any* principal axis with the positive $x$-axis we find

$$\tan 2\varphi = \frac{2F}{E - G} \quad . \tag{A11}$$

Although these equations get simpler if lensing properties are introduced (e. g. since $u_{(y)} = v_{(x)}$ the last equation reads $\tan 2\varphi = 2u_{(y)}/(u_{(x)} - v_{(y)})$) the complex notation is much more convenient. The equivalence of Eqs. (A10, A11) with Eqs.(12,13) follows after a little computation if the definition of $\mu$ Eq. (14) is fully expanded into normal derivatives

$$\mu = \frac{\left(u_{(x)} - v_{(y)}\right) + \mathrm{i}\left(v_{(x)} + u_{(y)}\right)}{\left(u_{(x)} + v_{(y)}\right) + \mathrm{i}\left(v_{(x)} - u_{(y)}\right)} \quad . \tag{A12}$$

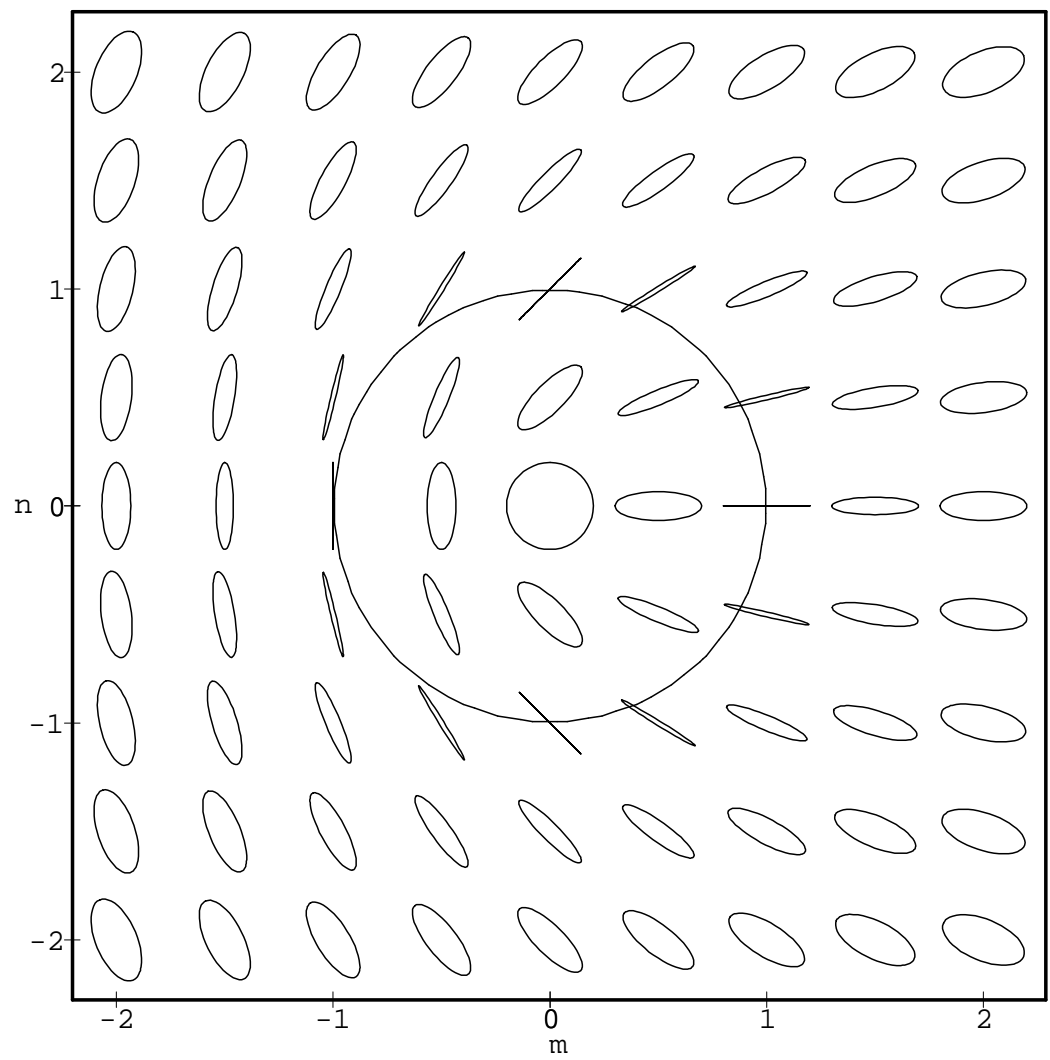

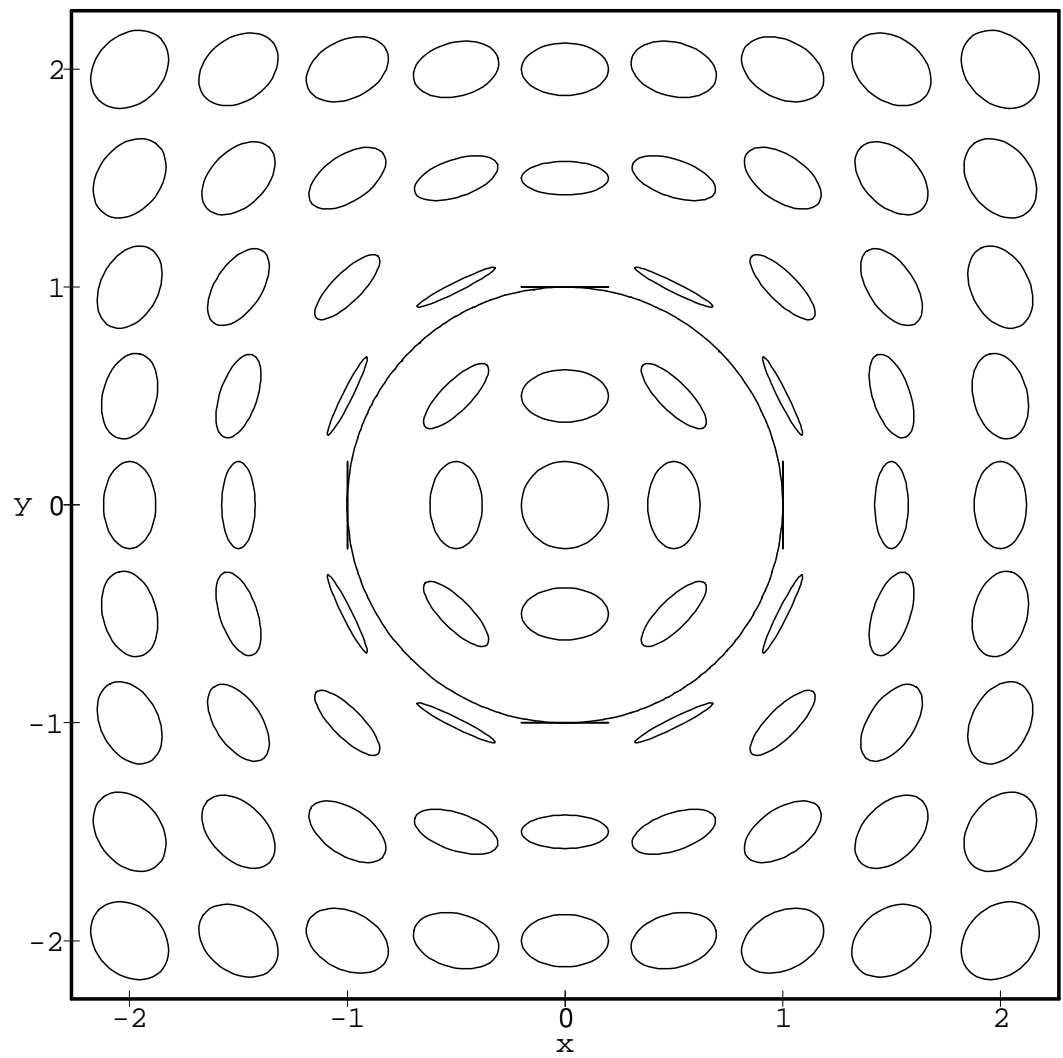

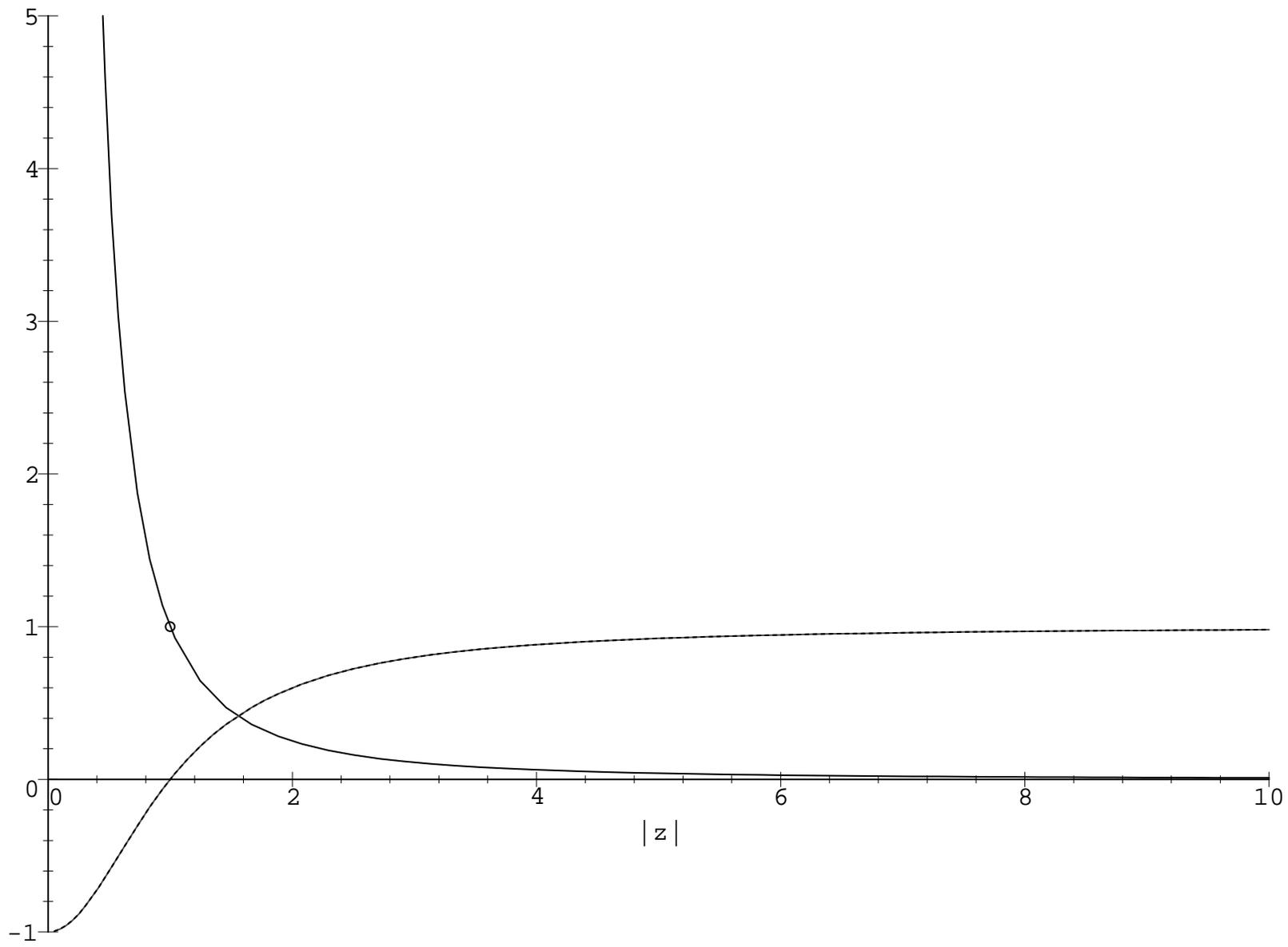

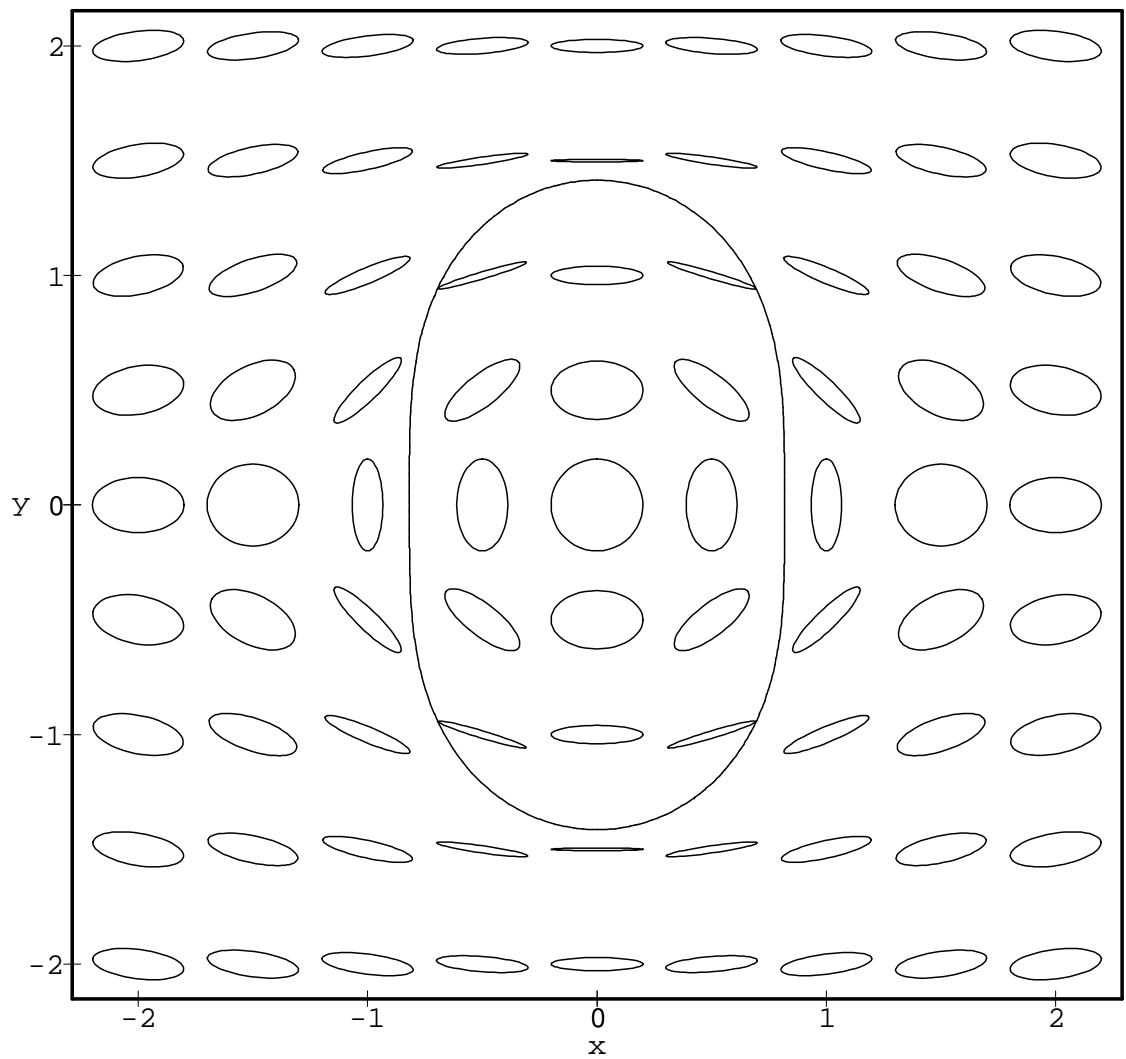

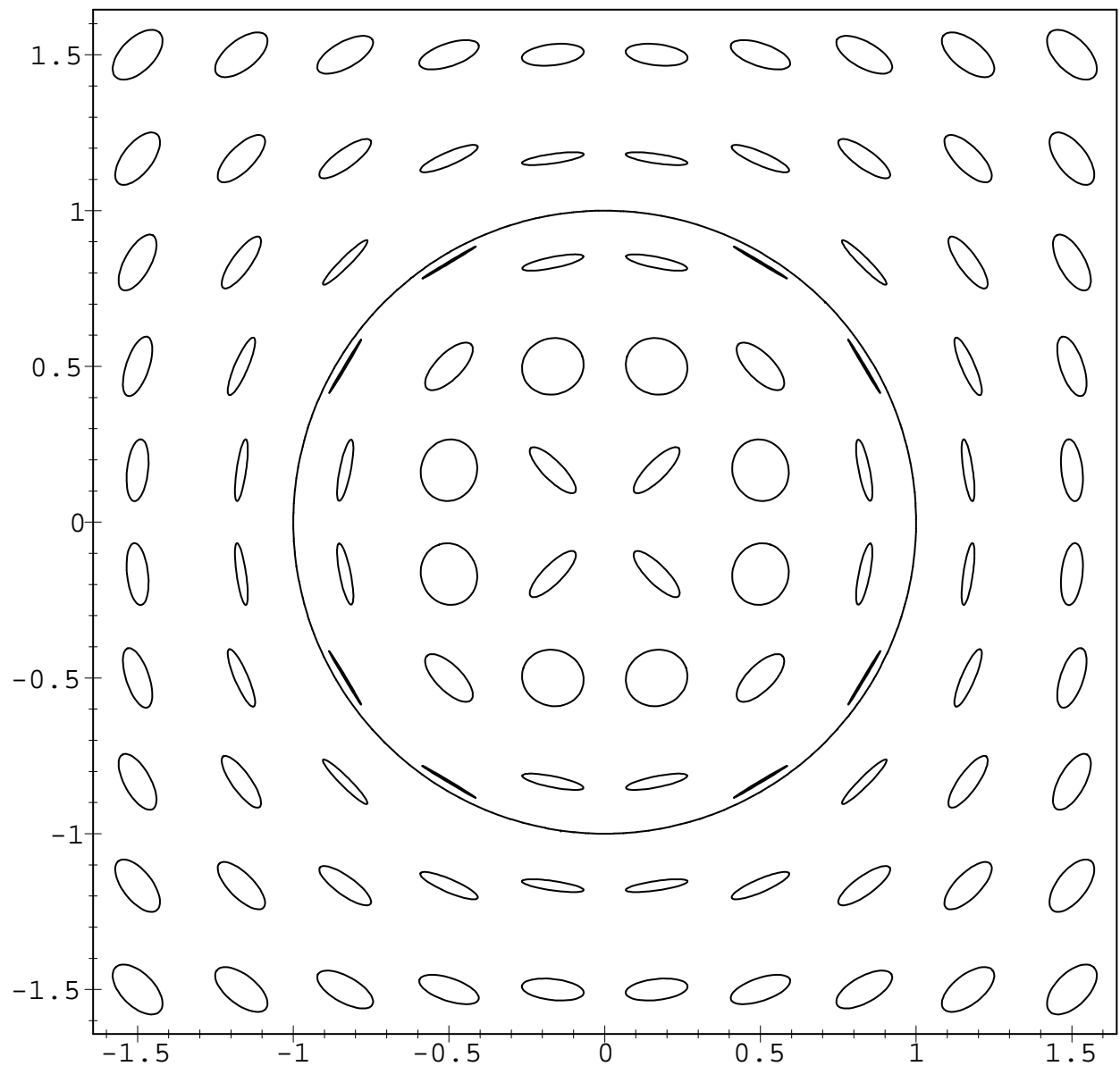

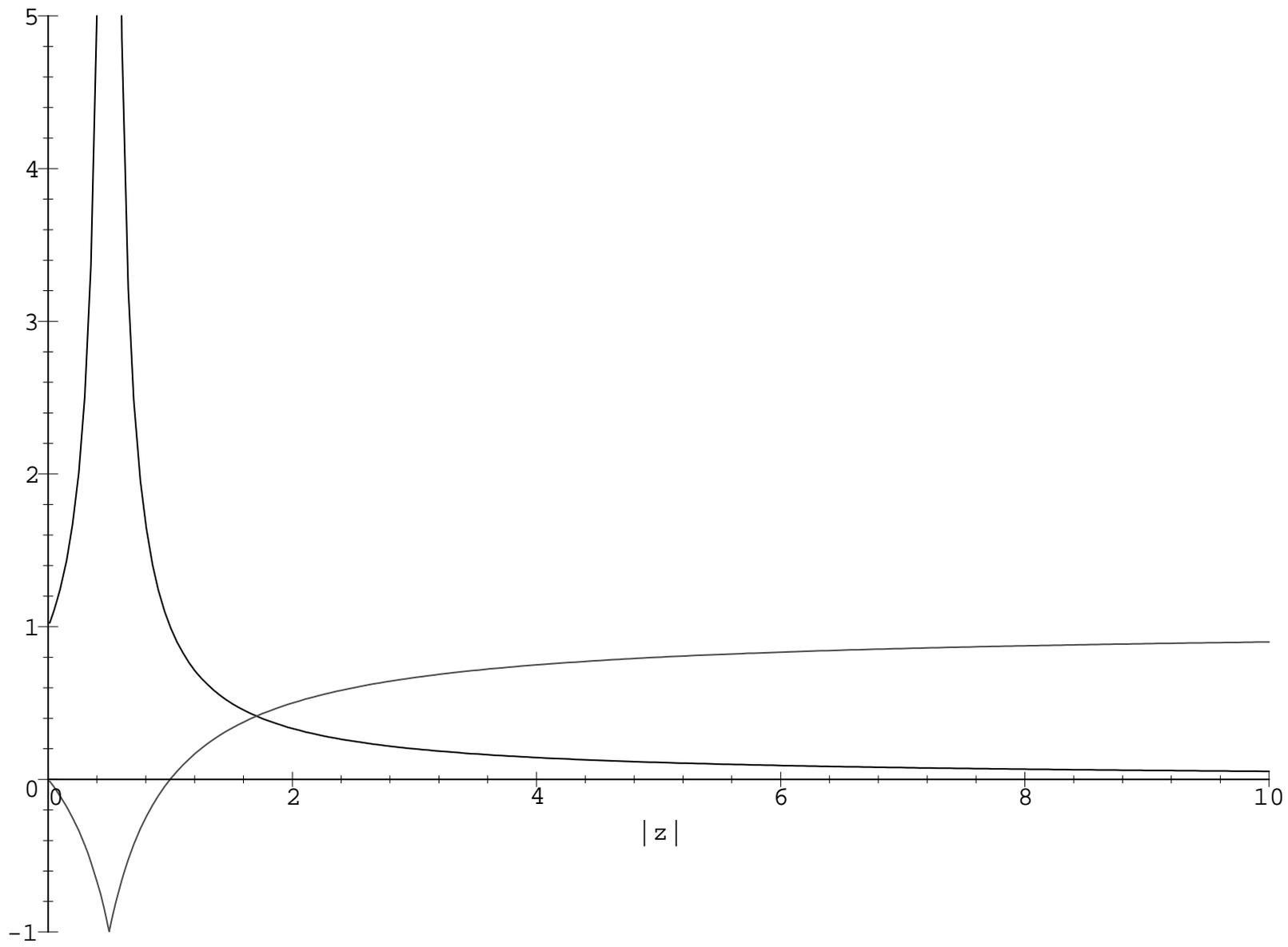

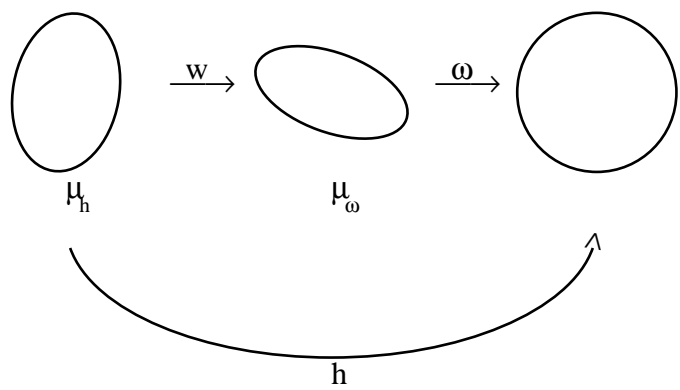